\documentclass[%
  twocolumn,
groupedaddress,
showpacs,preprintnumbers,
 amsmath,amssymb,
 aps,
 prd,
]{revtex4-1}

\usepackage{braket}            
\usepackage{graphicx}          
\usepackage[utf8]{inputenc}    
\usepackage{hyperref}          
\usepackage{dcolumn}           
\usepackage{color}             
\usepackage{adjustbox}         
\usepackage[caption=false]{subfig}        
\usepackage{csquotes}        
\usepackage{physics}

\usepackage{amsmath}

\makeatletter
\let\oldtheequation\theequation
\renewcommand\tagform@[1]{\maketag@@@{\ignorespaces#1\unskip\@@italiccorr}}
\renewcommand\theequation{(\oldtheequation)}
\makeatother

\newcommand{\mpi}{m_{\pi}}
\newcommand{\mpipacs}{m_{\pi}^{PACS}}

\newcommand{\sigmap}{$\Sigma^+$}
\newcommand{\sigmam}{$\Sigma^-$}
\newcommand{\cascadez}{$\Xi^0$}
\newcommand{\cascadem}{$\Xi^-$}

\begin{document}

\preprint{ADP-24-16-T1255}

\title{Quark mass effects in octet baryon magnetic polarisabilities via lattice QCD}
\author{Thomas Kabelitz}
\author{Waseem Kamleh}
\author{Derek Leinweber}

\affiliation{Special Research Centre for the Subatomic Structure of Matter (CSSM),\\
    Department of Physics, University of Adelaide, Adelaide, South Australia 5005, Australia
}

\date{\today}

\begin{abstract}
The quark mass dependence of octet baryon magnetic polarisabilities is examined at the level of
individual quark-sector contributions in the uniform background-field approach of lattice QCD.
The aim is to understand the direct impact of increasing the mass of a quark flavour on the
magnetic polarisability and indirect or environmental effects associated with changing the mass of
spectator quarks, insensitive to the background magnetic field.
Noting the need to set the electric charge of some quark flavours to zero,
a fractionally charged baryon formalism is introduced.
We find that increasing the mass of the charged quark flavour directly causes its contribution to
the magnetic polarisability to decrease.  However, increasing the mass of the spectator quark
flavour indirectly acts to increase the magnetic polarisability.
To gain a deeper understanding of these effects, we evaluate the predictions of the constituent
quark model in this context.  While the model provides a compelling explanation for the
environmental effect of varying the spectator quark mass, an explanation of the direct mass
dependence is more complicated as competing factors combine in the final result.  The lattice
results indicate the key factor is a reduction in the constituent quark magnetic moment with
increasing quark mass, as it governs the strength of the magnetic transition to the nearby decuplet
baryon.

\end{abstract}


\maketitle

\section{Introduction}

Understanding the manner in which individual quark sectors contribute to QCD observables provides a
deep understanding of QCD dynamics. The isolation of quark sectors arises
naturally in theoretical calculations as one considers the coupling of a probing current to each of
the quarks in turn.  By effectively setting the charge of all but one quark sector to zero, one can
disclose the contribution of a single quark sector.

With a quark sector isolated, one can then unravel the relationship between direct effects and
indirect effects due to changes in the environment in which the selected quark flavour resides.  For
example, consider the $u$-quark sector within the proton, isolated by setting the electric charge
of the $d$-quark to zero.  One can then probe the manner in which the $u$-sector contribution to a
proton electromagnetic observable changes as the $u$-quark mass is changed.  Herein, we will refer
to such effects as a "direct" effect.  However, "indirect" effects associated with changes
in the $d$-quark mass are also observable.  We refer to such indirect effects as
"environmental" effects, as the environment in which the $u$-quark resides changes.

Direct and indirect environmental effects in the electromagnetic form factors of octet baryons
\cite{Leinweber:1990dv}, decuplet baryons \cite{Leinweber:1992hy} and their transitions
\cite{Leinweber:1992pv}, were discovered long ago.  For magnetic moments, the discovery of
environmental effects was surprising as a basic tenant of the constituent quark model is a
constituent quark has an intrinsic magnetic moment governed solely by its mass.  There is no scope
for an environmental dependence.  However, such a dependence could explain violations in magnetic
moment sum rules \cite{Leinweber:1991vc} and it is central to early precise predictions of the
strange quark contribution to the proton's magnetic moment \cite{Leinweber:2004:precise},
predictions that have withstood the tests of modern determinations more than a decade later
\cite{Djukanovic:2019jtp,Alexandrou:2019olr}.

In this article we present the first examination of quark-sector contributions to octet baryon
magnetic polarisabilities.  We explore both direct and indirect environmental effects through the
variation of the masses of the quark flavours under consideration. Noting the need to set the
electric charge of some quark flavours to zero, a fractionally charged baryon formalism is
introduced to the background-field formalism.

In the process of understanding the underlying mechanisms that may be responsible to the effects
observed in our lattice QCD calculations, we refer to the constituent quark model
\cite{Close:1979bt}, improved to approximate recent lattice QCD results for octet baryons
\cite{kabelitz:octet}.

There the model highlights the importance of the $u$ quark in generating a large magnetic
polarisability and the impact of the octet-decuplet mass difference which suppresses the nucleon
magnetic polarisabilities relative to the hyperons, especially the $\Xi^0$. However, there are
more subtle discrepancies that remain to be understood.  For example, the subtle differences that
order the $p$, $n$, and \sigmap\ magnetic polarisabilities are not fully captured in the simple
model.

In this work, we aim to understand the direct and indirect effects in quark mass changes, to
elucidate the role of environmental effects in QCD \cite{Gross:2022hyw}. We aim to understand the
extent to which the current simple quark model contains environmental effects and explore a
possible role for additional effects that are yet to be considered in the model.

Our approach is via lattice QCD, building upon the established background field formalism.  We will
build on the advanced lattice techniques established in Ref.~\cite{kabelitz:octet}, this time
neutralising quark sectors to access the contributions one flavour at a time.  We will carefully
examine the direct and indirect effects through both lattice QCD and the constituent quark model,
allowing a determination of the capability of the constituent quark model and the opportunity to
evaluate whether environmental effects are as relevant for the magnetic polarisability as they are
for the magnetic moment.

In \autoref{sec:fractionallychargedbaryons} we define a fractionally charged baryon formalism that
enables an examination of direct and indirect environment effects.
In \autoref{sec:quarkmodel} we briefly revisit the magnetic polarisability in the constituent quark
model before we consider its predictions for fractionally charged baryons.
In \autoref{sec:backgroundfieldmethod} we review our implementation of the background field method.
The extraction of the magnetic polarisability on the lattice is presented in \autoref{sec:magneticpolarisability}.
In \autoref{sec:fitting} we discuss our fitting methodology and treatment of excited state
contamination of the underlying correlation functions required to extract the magnetic
polarisability in the background field method.
The details of the lattice simulation are presented in \autoref{sec:simulationdetails}.
In \autoref{sec:latticeresults} we present our results of the lattice simulation and evaluate the quark model's ability to describe them.
In \autoref{sec:conclusion} we summarise our findings.

\section{Fractionally charged baryons}\label{sec:fractionallychargedbaryons}

We begin by defining a formalism in which we can probe the direct and indirect environmental
effects of quark mass changes in the magnetic polarisabilities of octet baryons.  As in
Ref.~\cite{kabelitz:octet}, we consider the outer members of the baryon octet to avoid issues with
the mixing of $\Sigma^0$ and $\Lambda$ baryons in the background field.  As such, the quantum
numbers of these baryons are carried by two quark flavours. We are specifically interested in two
concepts.

\emph{Direct quark mass dependence}: How the contribution of a quark sector changes as its mass
changes between that of a down quark and that of a strange quark ({\it e.g.} $d\rightarrow s$ in
$n\rightarrow\Xi^0$).

\emph{Indirect environmental dependence}: How the contribution of a quark sector changes as the
mass of the other sector changes between that of a down and strange quark ({\it e.g.} $u$ in
$p\rightarrow\Sigma^+$).

\begin{table}[tb]
\caption{The properties of the four quark flavours considered in isolating individual quark-sector
  contributions to baryon magnetic polarisabilities.}
\label{tab:flavourDefs}
    \begin{ruledtabular}
    \begin{tabular}{cccc}
    \noalign{\smallskip}
    Quark   & Symbol & Electric Charge & Mass  \\
    \noalign{\smallskip}
    \hline
    \noalign{\smallskip}
    down    & $d$    & $-1/3$ & $m_d$ \\
    strange & $s$    & $-1/3$ & $m_s$ \\
    light   & $l$    & 0      & $m_d$ \\
    heavy   & $h$    & 0      & $m_s$ \\
    \end{tabular}
    \end{ruledtabular}
\end{table}

We determine these contributions by constructing baryons with a single electrically charged flavour
and a neutrally charged flavour, thereby isolating the electrically charged flavour's
contribution. To do so we utilise the four quark flavours defined in Table~\ref{tab:flavourDefs}
and construct eight baryons with the flavours
\begin{center}
\begin{tabular}{llll}
    $ddl$,\qquad\null & $lld$,\qquad\null & $ssl$,\qquad\null & $lls$, \\
    $ddh$, & $hhd$, & $ssh$, & $hhs$.
\end{tabular}
\end{center}

These eight baryons allow us to examine the direct quark mass dependence
\begin{align}
    ddl & \rightarrow ssl, \nonumber \\
    ddh & \rightarrow ssh, \nonumber \\
    lld & \rightarrow lls, \nonumber \\
    hhd & \rightarrow hhs, \label{eqn:4direct}
\end{align}
and the indirect environment dependence
\begin{align}
    ddl & \rightarrow ddh, \nonumber \\
    ssl & \rightarrow ssh, \nonumber \\
    lld & \rightarrow hhd, \nonumber \\
    lls & \rightarrow hhs, \label{eqn:4indirect}
\end{align}
of all possible charge and environment sectors. To examine the direct and environmental dependence,
we examine differences in the magnetic polarisability such as $\beta_{ssl} - \beta_{ddl}$ for the
direct mass effect. In calculating the difference, we will take advantage of correlations between
the quantities to accurately discern changes due to variation of the quark masses.

\section{Constituent quark model}\label{sec:quarkmodel}

In a simple constituent quark model, the constituent quark masses are tuned to reproduce the
magnetic moments of the $p$, $n$, and $\Lambda$ baryons \cite{pdg2020}. Such a simple model relies
on the assumption that the quarks are blind to their environment.  For example, this assumption
makes the contribution of the $u$ quark the same in the $n\, (ddu)$ and the $\Xi^0\, (ssu)$, despite the
change in the mass of its environment. However, experimental evidence of environmental effects was
highlighted in Ref.~\cite{Leinweber:2004:precise} for the magnetic moments. Nonetheless, the
constituent quark model retains significant value. Its ability to simply describe the qualitative
pattern exhibited by the baryon magnetic moments is complemented by a remarkable ability to explain
the broad behaviours of the magnetic polarisability of the octet baryons as presented in
Ref.~\cite{kabelitz:octet}.

In that work we derived the following expression for the magnetic polarisability of an outer octet
baryon 
\begin{align}\label{eqn:quarkmodel:polarisability}
    \beta &=
		\frac{1}{2\pi}\frac{\abs{\bra{\mathcal{B}}\hat{\mu}_z\ket{\mathcal{B}^*}}^2}{m_{\mathcal{B}^*}-m_{\mathcal{B}}}
        - \sum_{f=1}^3\frac{q_f^2\,\alpha}{6 m_f}\expval{r^2}_f \, , \nonumber \\
        &=
		\frac{4}{9\pi}\frac{\abs{\mu_D-\mu_S}^2}{m_{\mathcal{B}^*}-m_{\mathcal{B}}}
        - \sum_{f=1}^3\frac{q_f^2\,\alpha}{6 m_f}\expval{r^2}_f \, , \nonumber \\
    &\equiv \beta_1 - \beta_2 \, ,
\end{align}
where $\alpha={e^2}/{4\pi}$ is the fine structure constant. $\beta_1$ depends on the magnetic
transition moment between octet and decuplet baryons for the outer members of the baryon octet
\begin{equation}
    \bra{\mathcal{B}}\hat{\mu}_z\ket{\mathcal{B}^*} = \frac{2\sqrt{2}}{3}\left(\mu_D - \mu_S\right) \, ,
\end{equation}
expressed in terms of the constituent quark moments of the doubly, $\mu_D$, and singly, $\mu_S$,
represented sectors for an outer octet baryon. In the constituent quark model, the constituent
quark moments are given in terms of the constituent quark masses, $m_f$, for each flavour, $f$
\begin{equation}\label{eqn:quarkmodel:quarkmoment}
    \mu_f = \frac{q_f\, e}{2\, m_f} \, .
\end{equation}
The contributions of opposing sign in Eq.~\ref{eqn:quarkmodel:polarisability} produce interesting
complexity \cite{kabelitz:octet}. $\beta_1$ probes the transition to an intermediate excitation
through virtual photon exchange, while $\beta_2$ probes the distribution of quarks within the
hadron.

\subsection{Implementing the quark model}

The quark model predictions require the model's input quantities be determined on the lattice and
interpolated. The details of this interpolation are described in depth in
Ref.~\cite{kabelitz:octet}.
To estimate the constituent quark masses away from the physical point, we utilise a simple
model
\begin{align}\label{eqn:quarkmodelimplementation:octetmassmodel}
    m^0_{\rm oct} + \alpha_{\rm oct}\,(3\,m_l) &= m_N, \nonumber\\
    m^0_{\rm oct} + \alpha_{\rm oct}\,(m_l+2\,m_h) &= m_{\Xi}.
\end{align}
Here $m_l=m_d$ and $m_h=m_s$ are the light and heavy constituent quark masses already determined in
the simple constituent quark model via baryon magnetic moments at the physical point, as summarised
in Ref.~\cite{pdg2020}.  $m^0_{\rm oct}$ and $\alpha_{\rm oct}$ are fit parameters, constrained by
the physical nucleon and $\Xi$ masses.  With the parameters determined, one can then use
Eq.~\ref{eqn:quarkmodelimplementation:octetmassmodel} with baryon masses from our lattice QCD
simulations to solve for the constituent quark masses, $m_l$ and $m_h$, away from the physical
point.

We linearly interpolate the octet baryon masses as a
function of pion mass~\cite{Walker-Loud:2014:ruler}, allowing the determination of the constituent
quark masses as a function of the pion mass.

In the spirit of the constituent quark model, we utilise these constituent quark masses and
equations of the form Eq.~\ref{eqn:quarkmodelimplementation:octetmassmodel} to determine the octet
baryon masses, including the masses of fractionally charged baryons that would
otherwise not exist. This includes the mass of an octet $ssh$ or $hhs$ with a mass of
\begin{equation}
    m_{ssh} = m_{hhs} = m^0_{\rm oct} + \alpha_{\rm oct}\,(3\,m_h)\, .
\end{equation}
The decuplet baryon masses are determined using an equivalent model with fit parameters determined
in the same manner using the same physical point $m_l$ and $m_h$ constituent quark masses, and the
physical $\Delta$ and $\Omega$ masses.

Finally, we interpolate the quark distribution radii determined on the PACS-CS ensembles by Stokes
{\it et al.}~\cite{stokes2020formfactor}. The proton and neutron squared radii are characterised
well with an interpolation that is linear in $\log(\mpi)$. Quark distribution radii for single quark
flavours of unit charge in the doubly $(D)$, and singly $(S)$, represented sectors are defined by
\begin{align}
    \expval{r^2}_p &= \phantom{-} 2\, \frac{2}{3}\, \expval{r^2}_D - \frac{1}{3} \expval{r^2}_S\, , \\
    \expval{r^2}_n &=            -2\, \frac{1}{3}\, \expval{r^2}_D + \frac{2}{3} \expval{r^2}_S\, ,
\end{align}
where charge and quark-counting factors are explicit.  The resulting distribution radii for single
quarks of unit charge are
\begin{align}
    \expval{r^2}_D &= \frac{1}{2}\left(2\expval{r^2}_p + \expval{r^2}_n\right), \\
    \expval{r^2}_S &= \expval{r^2}_p + 2\expval{r^2}_n.
\end{align}
As only the proton and neutron were examined in Ref.~\cite{stokes2020formfactor}, we approximate
the doubly and singly represented strange-quark distribution radii to be that of the light quark at
the heaviest PACS-CS quark mass labelled by $m_\pi = 701$ MeV.

\subsection{Constituent quark model predictions for the magnetic polarisability}

\begin{figure}
    \includegraphics[width=0.45\textwidth]{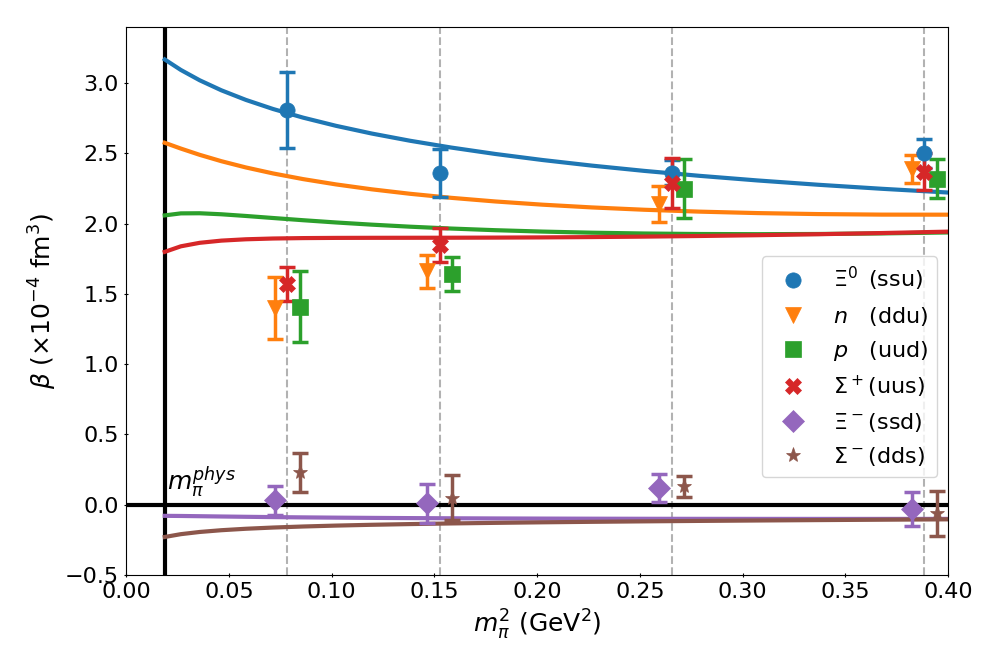}
    \caption{Improved 
      quark model predictions from Ref.~\cite{kabelitz:octet} for the magnetic
      polarisabilities of octet baryons (curves) are compared with lattice QCD determinations
      (points) calculated on the PACS-CS ensembles considered herein.  The legend is ordered to
      match the vertical ordering of the quark model curves at the physical pion mass. Dashed lines
      represent the pion masses of the PACS-CS ensembles.}
    \label{fig:quarkmodel:polarisabilitypredictions}
\end{figure}

Using the above implementation, the magnetic polarisability of the outer octet baryons are
predicted in Ref.~\cite{kabelitz:octet}. A simple magnitude correction is required
\begin{equation}
    \beta = a_1\beta_1 - a_2\beta_2 \, ,
\end{equation}
where the fit parameters $a_1=0.401,\, a_2=0.532$ best reproduce the lattice QCD data. The
phenomenology associated with this scaling is discussed in Ref.~\cite{kabelitz:octet}.  With
this improvement, the quark model is able to capture the qualitative features of the lattice QCD
determinations as illustrated in Fig.~\ref{fig:quarkmodel:polarisabilitypredictions}.

Specifically, it predicts the small nature of the magnetic polarisability of the \sigmam\ and
\cascadem\ baryons, which was otherwise not known. It also correctly predicts the large magnetic
polarisability of the \cascadez. Finally, it predicts approximately the correct magnitude for the
magnetic polarisability of the $p$, $n$, and \sigmap\ though there is a lack of clarity about
their ordering. The successes of the model are significant given its simplicity.

The model also highlights the importance of the $u$ quark in generating a large magnetic
polarisability through the transition term $\beta_1$ with the factor $(\mu_D -
\mu_S)^2$. Combinations of $\mu_u$ and $\mu_d$ or $\mu_s$ add and square to a large magnitude,
whereas a combination of $\mu_d$ and $\mu_s$ approximately cancel and become small in the square.
This latter point explains the small magnitude of the \sigmam\ and \cascadem\ magnetic
polarisabilities.

The more subtle discrepancies between the model and the lattice simulation results may be
associated with an environmental effect in the constituent quark moments as discussed in the
introduction.  Thus, we turn our attention to the contributions of individual quark sectors with an
aim to understanding direct and indirect environmental effects governing their contributions to the
magnetic polarisability.

\subsection{Fractionally charged baryon formalism}\label{sec:quarkmodel:fractionallychargedbaryons}

We first evaluate the quark model predictions when considering fractionally charged
baryons. Consider the magnetic polarisability of Eq.~\ref{eqn:quarkmodel:polarisability}
for the proton
\begin{align}
    \beta_{uud} & =
    \frac{4}{9\pi}\frac{\abs{\mu_u-\mu_d}^2}{m_{\Delta}-m_N}
    - \frac{8}{9}\frac{\alpha}{6 \, m_l}\expval{r^2}_l-\frac{1}{9}\frac{\alpha}{6 m_l}\expval{r^2}_l \, , \nonumber \\
                & =
    \frac{4}{9\pi}\frac{9\mu_d^2}{m_{\Delta}-m_N}
    - \frac{\alpha}{6 \, m_l}\expval{r^2}_l \, ,
\end{align}
where we have used the charge symmetric relation \mbox{$\mu_u = -2\mu_d$}.  Now consider neutralising the
$u$ quark sector for the $lld$ baryon
\begin{align}
    \beta_{lld} & =
    \frac{4}{9\pi}\frac{\abs{-\mu_d}^2}{m_{\Delta}-m_N}
    -\frac{1}{9}\frac{\alpha}{6 \, m_l}\expval{r^2}_l \, , \nonumber \\
                & =
    \frac{4}{9\pi}\frac{\mu_d^2}{m_{\Delta}-m_N}
    - \frac{1}{9}\frac{\alpha}{6 \, m_l}\expval{r^2}_l \, .
\end{align}
We note a significant drop in the magnitude by a factor of nine. Our focus on the elemental $d$
quark in the lattice QCD simulations means that this reduction by an order of magnitude will have
significant implications that will be discussed in due course.

\subsubsection{Environmental dependence in the quark model}
\label{sec:QMindirect}

Considering the environmental dependence in $lld\rightarrow hhd$, we note that the charged sector is unchanged. As a result, when taking the difference
\begin{equation}
    \beta_{hhd} - \beta_{lld} = \frac{4}{9 \, \pi}\left[\frac{\mu_d^2}{m_{\Xi^*} - m_{\Xi}} - \frac{\mu_d^2}{m_{\Delta} - m_N}\right],
\label{eqn:hhd-lld}
\end{equation}
the quark distribution term vanishes and we are left only with the difference in the octet-decuplet mass splittings.
The quark model's prediction of a nontrivial environmental effect for the magnetic polarisability
contrasts that for the magnetic moment.

The result of Eq.~\ref{eqn:hhd-lld} can be easily extended to the other three environmental effects
of Eq.~\ref{eqn:4indirect}.  In every case, there are additional heavy quarks in the right-hand
baryon of Eq.~\ref{eqn:4indirect}.  As hyperfine splittings are inversely proportional to the
product of the constituent quark masses, the octet-decuplet mass splitting is smaller for baryons
with more heavy quarks.  Thus, the decuplet $-$ octet baryon mass splitting is smaller for the
leading term and the sign of all environment effects are predicted to
be positive.

\subsubsection{Direct quark mass dependence in the quark model}
\label{sec:QMdirect}

In the case of the mass dependence, the mass of the charged sector is changing and the expression
becomes more complex. Considering $lld \rightarrow lls$
\begin{align}
    \beta_{lls} - \beta_{lld} =
      \frac{4}{9\,\pi}&\left[\frac{\mu_s^2}{m_{\Sigma^*} - m_{\Sigma}} - \frac{\mu_d^2}{m_{\Delta}
        - m_N}\right] \nonumber \\
    - & \left[\frac{\alpha \, q_s^2}{6 \, m_h}\expval{r^2}_h - \frac{\alpha \, q_d^2}{6 \,
        m_l}\expval{r^2}_l\right] \, ,
    \label{eqn:lls-lld}
\end{align}
we observe none of the previous cancellation.  Instead we observe competing effects.

As $\mu_d^2 > \mu_s^2$ and $(m_{\Delta} - m_N)^{-1} < (m_{\Sigma^*} - m_{\Sigma})^{-1}$
the sign of the first term depends on the interplay between these two factors.  However the sign of the
second term is clear as $\expval{r^2}_h/m_h < \expval{r^2}_l/m_l$ and thus the contribution
including the leading minus sign is positive. Therefore, if $\beta_{lls} - \beta_{lld} < 0$ then one
can point to $\mu_d^2 > \mu_s^2$ as the dominant effect.

Similar arguments hold for the case of $hhd \to hhs$.  If $\beta_{hhs} - \beta_{hhd} < 0$ then one
can point to $\mu_d^2 > \mu_s^2$ as the dominant effect.

Now consider $ddl \rightarrow ssl$,
\begin{align}
    \beta_{ssl} - \beta_{ddl} =
      \frac{4}{9\,\pi}&\left[\frac{\mu_s^2}{m_{\Xi^*} - m_{\Xi}} - \frac{\mu_d^2}{m_{\Delta} - m_N}\right] \nonumber \\
    - \, 2\, &\left[\frac{\alpha \, q_s^2}{6 \, m_h}\expval{r^2}_h - \frac{\alpha \, q_d^2}{6 \,
        m_l}\expval{r^2}_l\right]\, .
    \label{eqn:ssl-ddl}
\end{align}
This time the octet-decuplet mass splitting creates even greater tension between competing effects
as does the charge distribution term.
While $\mu_d^2 > \mu_s^2$ as before, $(m_{\Delta} - m_N)^{-1} \ll (m_{\Xi^*} - m_{\Xi})^{-1}$ and thus
there is greater competition between the factors in setting the sign. The sign of the
second term is as before, but with double the magnitude.

If $\beta_{ssl} - \beta_{ddl} < 0$ then one
can once again point to $\mu_d^2 > \mu_s^2$ as the dominant effect.  However, the quark model
predicts a suppressed effect relative to $\beta_{lls} - \beta_{lld}$ due to a reduced difference in
mass splittings and the magnitude of the quark distribution term.

Finally, similar arguments hold for the case of $ddh \to ssh$.  If $\beta_{ssh} - \beta_{ddh} <
0$ then one can point to $\mu_d^2 > \mu_s^2$ as the dominant effect.  Again the effect is predicted
to be suppressed relative to $hhd \to hhs$.

To summarise, if the change in the constituent quark magnetic moment dominates the direct
quark-mass effect, the case of $hhd \to hhs$ holds the greatest promise for illustrating an effect
in the lattice QCD simulations. The prevalence of heavy quarks will aid in suppressing statistical
uncertainties and the tension between $\mu_d^2 > \mu_s^2$ and competing effects is lowest for
$\beta_{hhs} - \beta_{hhd}$.  If $\beta_{hhs} - \beta_{hhd} < 0$ then one has a simple intuitive
picture of the physics driving the direct quark mass effect.  It is the simple suppression of the
constituent quark moment that suppresses transitions to nearby decuplet baryons and thus diminishes
the polarisability.

We now proceed to the lattice QCD calculation to provide reference to these predictions.

\section{Background Field Implementation}\label{sec:backgroundfieldmethod}

The background field method is a well established approach to calculating the magnetic
polarisability in a lattice QCD calculation. The approach used in this work is identical to our
approach in Ref.~\cite{kabelitz:octet}. Full details are given there.

In the background field method, a minimal coupling to the discretised covariant derivative results
in the application of a phase to the QCD gauge field. Through the requirement for periodicity at
the boundary, we obtain a quantisation condition for the magnetic field strength $B$ given by
\begin{equation}\label{eqn:backgroundfield:quantisationcondition}
    e\,B=\frac{2\,\pi}{N_x\,N_y\,a^2}\frac{1}{q_d}\,k_d,
\end{equation}
where $k_d$ is an integer specifying the field strength in multiples of the minimum uniform field
strength quantum. The condition is written in terms of the down quark charge $q_d\,e$, the number
of spatial sites $N_x,\,N_y$ and the lattice spacing $a$.

The magnetic field is defined such that the field corresponding to integer $k_d=1$ is oriented in
the negative $\hat{z}$-direction. This work utilises quark propagators and baryon correlation
functions calculated at \mbox{$k_d=0,\,\pm 1,\, \pm 2$}. We will refer to the strength of the magnetic
field in terms of $k_d$ throughout this work.

\section{Lattice QCD Formalism}\label{sec:magneticpolarisability}

In the presence of a uniform background magnetic field, the energy of a baryon changes as a
function of magnetic field strength \cite{Martinelli:1982:expansion,primer2014magnetic}
\begin{equation}\label{eqn:backgroundfield:energyexpansion}
    E(B) = m + \vec{\mu}\cdot\vec{B} + \frac{|q_B\,e\,B|}{2\,m}\left(n+1\right) -
    \frac{1}{2}\, 4 \pi\,\beta\,B^2 + \order{B^3} \, .
\end{equation}
Here the mass of the baryon, $m$, is complemented by contributions from the magnetic moment
$\vec{\mu}$, the Landau term proportional to $|q_B\, e\, B|$ where $q_B$ is the charge of the
baryon, and the magnetic polarisability $\beta$. As the baryons in this work are fractionally
charged, we use a U(1) Landau-mode projection to select $n=0$. This aspect of the calculation is
detailed in \autoref{sec:simulationdetails:landauprojection}.

As we demonstrated in Ref.~\cite{kabelitz:octet}, the contribution from the magnetic moment may be
subtracted by consideration of the average of "spin-field aligned" and "spin-field anti-aligned"
energy. The baryon mass may also be removed through the subtraction of the zero-field
energy.

This process may be mirrored in a correlation function ratio which enables cancellation of
correlated fluctuations. We define a "spin-field aligned" correlator
\begin{equation}\label{eqn:polarisability:alignedcorrelator}
    G_{\uparrow\uparrow}(B) = G(+s,+B) + G(-s, -B) \, ,
\end{equation}
where the baryon's spin is aligned with the magnetic field and a "spin-field anti-aligned"
correlator
\begin{equation}\label{eqn:polarisability:antialignedcorrelator}
    G_{\uparrow\downarrow}(B) = G(+s,-B) + G(-s, +B) \, ,
\end{equation}
where the spin and field are opposed. These correlators are combined in the ratio
\begin{equation} \label{eqn:polarisability:ratio}
    R(B,t) = \frac{G_{\uparrow\uparrow}(B,t)\,G_{\uparrow\downarrow}(B,t)}{G(0,t)^2} \, ,
\end{equation}
which aggregates the positive and negative field strengths together. Due to this aggregation, any
reference to the magnetic field strength from this point refers to the aggregated positive field
strength.

The effective energy of this correlation function ratio acts to subtract the magnetic moment term
and baryon mass. As such, we define the magnetic polarisability energy shift $\delta
E_{\beta}(B,t)$
\begin{align}\label{eqn:polarisability:energyshift}
    \delta E_{\beta}(B,t) & = \frac{1}{2}\frac{1}{\delta t}\lim_{t\rightarrow \infty}\log\Bigl(\frac{R(B,t)}{R(B,t+\delta t)}\Bigr),\nonumber \\
                          & = \frac{1}{2}\left[
        \delta E_{\uparrow\uparrow}(B) + \delta E_{\uparrow\downarrow}(B)
    \right] - \delta E(0)\nonumber                                                                                                            \\
                          & = \frac{\abs{q\, e\, B}}{2m} - \frac{4\pi}{2}\beta|B|^2 +\order*{B^3}.
\end{align}

However, we require more than the magnetic polarisability of a single baryon, instead we require
the difference in magnetic polarisability between two baryons. We saw in
\autoref{sec:fractionallychargedbaryons} that fractionally charged baryons have magnetic
polarisabilities an order of magnitude smaller than that of the physical baryons composed with a
$u$ quark. As such, extracting sufficient signal to take a meaningful difference is
difficult. Extracting the magnetic polarisability of the fractionally charged baryons separately
and determining a correlated difference often fails to produce meaningful insight, especially at
lighter quark masses.

However, the highly correlated nature of the two baryons can be exploited by calculating the
difference at the correlator level. As such, we construct the ratio $R(B,t)$ for both baryons in
question and construct the following effective energy shift
\begin{align} \label{eqn:polarisability:differenceenergyshift}
    \delta E_{\beta_1-\beta_2}(B,t) & = \frac{1}{2}\frac{1}{\delta t} \lim_{t\rightarrow\infty}\log\left(
    \frac{R_{\beta_1}(B,t)R_{\beta_2}(B,t + \delta t)}{R_{\beta_1}(B,t + \delta t)R_{\beta_2}(B,t)}
    \right), \nonumber \\
    & =\left(\frac{\abs{q_{B_1}}}{m_{B_1}}-\frac{\abs{q_{B_2}}}{m_{B_2}}\right)\frac{\abs{eB}}{2}
    - \frac{4\pi}{2}\left(\beta_1 - \beta_2 \right)\abs{B}^2 \nonumber \\
    & \phantom{=} + \order{B^3}.
\end{align}
This allows for fitting of the magnetic polarisability difference $\beta_1-\beta_2$ directly,
providing maximal cancellation of correlated fluctuations. We note that the magnetic polarisability
differences we are interested in here will always have $q_{B_1} = q_{B_2}$, hence the quark charge
may be factored out of the difference in the Landau term.

\section{Fitting} \label{sec:fitting}

To extract the magnetic polarisability difference $\beta_1 - \beta_2$ we fit
Eq.~\ref{eqn:polarisability:differenceenergyshift} as a function of field strength. In practice, it
is simpler to fit in terms of the field strength quanta $k_d$. Using the quantisation condition of
Eq.~\ref{eqn:backgroundfield:quantisationcondition}
\begin{equation}
    e\,B=\frac{2\,\pi}{N_x\,N_y\,a^2}\frac{1}{q_d}\,k_d,
\end{equation}
we substitute for $e\, B$ in Eq.~\ref{eqn:polarisability:differenceenergyshift} to re-write $\delta E_{\beta_1-\beta_2}$ in terms of the field strength quanta
\begin{equation}
  \delta E_{\beta_1-\beta_2}(k_d) = L(k_d, m_{B_1},m_{B_2}) + C \, \left(\beta_1-\beta_2\right) \, k_d^2 \, , \label{eqn:fitting:fieldstrengthfit}
\end{equation}
where
\begin{equation}
  L(k_d, m_{B_1},m_{B_2}) = \frac{1}{2}\frac{2\pi}{N_x\, N_y \, a^2}\left(\frac{\abs{q_{B_1}}}{m_{B_1}}-\frac{\abs{q_{B_2}}}{m_{B_2}}\right)\abs{\frac{1}{q_d} \, k_d},
\end{equation}
and
\begin{equation}
  C = - \frac{1}{2 \, \alpha}\left[\frac{2\pi}{N_x \, N_y \, a^2}\right]^2\frac{1}{q_d^2}.
\end{equation}

It is important to note that $q_d$ is the down quark charge while $q_{B_1},q_{B_2}$ are the
hadronic charges of the two baryons.  For convenient future reference, we have defined
$L(k_d,m_1,m_2)$ the combined Landau contribution and $C$ the remaining coefficient to the magnetic
polarisability difference.

We fit $\delta E_{\beta_1-\beta_2}$ at two positive field strengths to extract the magnetic
polarisability difference $\beta_1-\beta_2$.
We obtain the best estimate for $\delta E_{\beta_1-\beta_2}(k_d)$ through a weighted average of fit
windows which provides a systematic approach to plateau fitting \cite{NPLQCD:2020multihadron}. The
method is detailed fully in Ref.~\cite{kabelitz:octet}.

In the method, the $i$th eligible window of $N$ candidates is assigned a weight
according to
\begin{equation}\label{eqn:fitting:weights}
    w_i = \frac{1}{\mathcal{Z}}\frac{p_i}{(\delta E_i)^2}\, ,
\end{equation}
where
\begin{equation}
    \mathcal{Z} = \sum_{i=1}^N\frac{p_i}{(\delta E_i)^2}\, .
\end{equation}
Here, $\delta E_i$ is the uncertainty of the $i$th fit, and $p_i$ is the $p$-value of the fit. The
$p$-value is the probability of a given fit occurring in a $\chi^2$ distribution.
With these weights, the average effective energy is
\begin{equation}
    E = \sum_i^n w_i \, E_i \, ,
\end{equation}
and the statistical error
\begin{equation}
    (\delta E)^2 = \sum_i^n w_i(\delta E_i)^2\, .
\end{equation}

Eligible windows must satisfy a few criteria. To avoid fitting noise, we define a $t_{\rm max}$,
the last time slice before signal is clearly lost to noise. We define a corresponding $t_{\rm
  min}$, the first time slice at which single state isolation has been reached for all underlying
(spin-field aligned, anti-aligned, zero-field) correlation functions. We also require the fit
window to be of minimum length three and $t_f = t_{\rm max}$. This fixing of the endpoint ensures
that early time-slices are not disproportionately favoured.

In our examination of the octet baryons in Ref.~\cite{kabelitz:octet} we found it vital to impose
the condition of single state isolation on all underlying correlation functions when determining
$t_{\rm min}$. In constructing the correlation function ratio Eq.~\ref{eqn:polarisability:ratio}
(which enables the extraction of the magnetic polarisability) the resulting correlator can display
plateau-like behaviour before the underlying correlation functions have reached single state
isolation.

\begin{figure*}
    \includegraphics[width=0.75\linewidth]{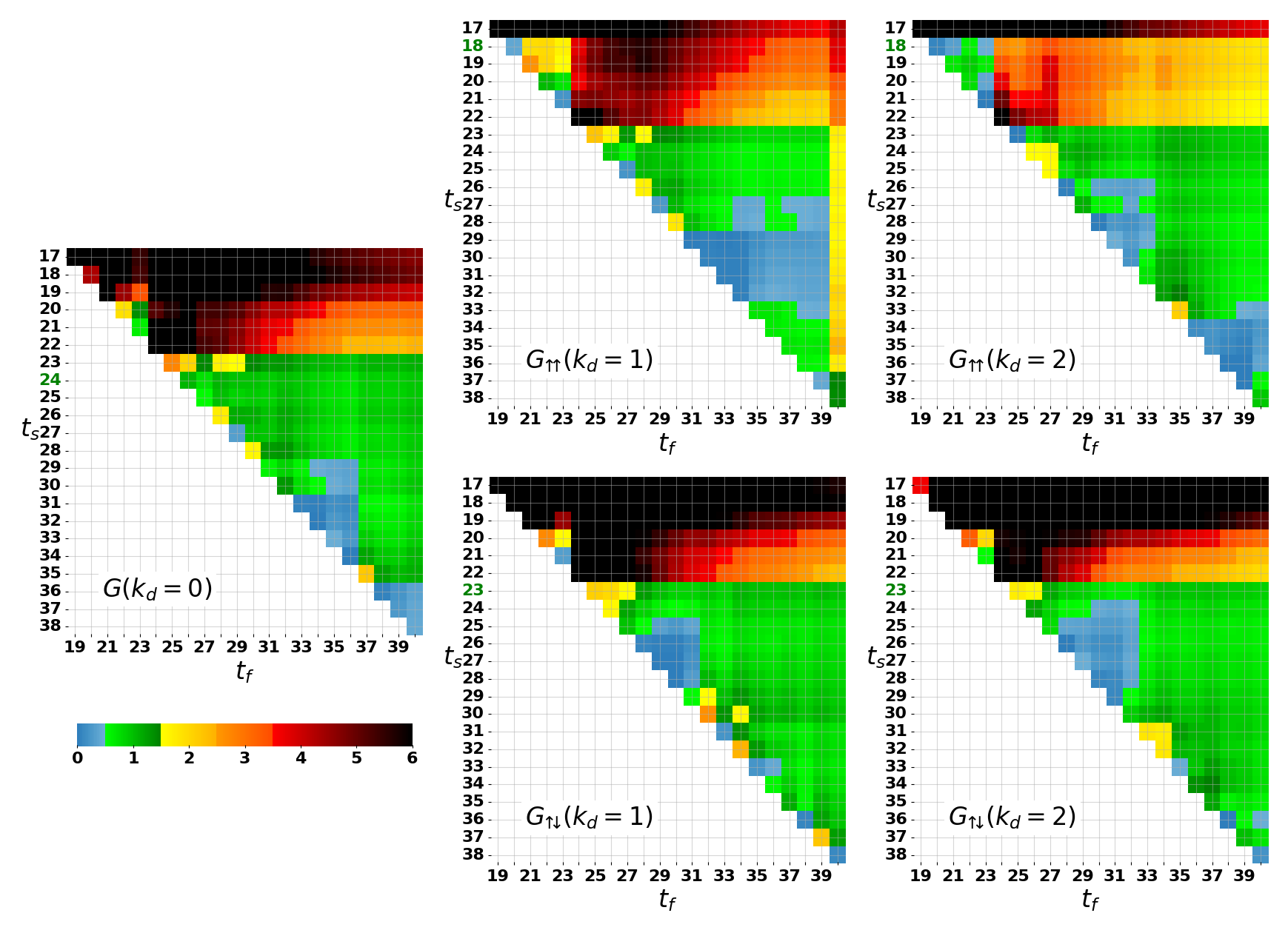}
    \caption{$\chi^2_{\rm dof}$ heat maps for fits to $\log G$ for all fit windows from $t_s
      \rightarrow t_f$ of the aligned ($\uparrow\uparrow$) and anti-aligned ($\uparrow\downarrow$)
      correlators at each field strength of interest. The numbers in parentheses denote the field
      strength quanta governed by $k_d$. This example is a $ddl$ baryon at $\kappa=0.13700$,
      $\mpi=296\,$MeV. The first time slice at which single-state isolation has been reached is
      highlighted green.}
    \label{fig:fitting:ddlheatmaps}
\end{figure*}

As we did in that work, we utilise the $\chi^2_{\rm dof}$ of a linear fit to $\log G$ as a metric
for excited state contamination. An easy interface to the large number of $\chi^2_{\rm dof}$ to be
checked are the set of heatmaps we construct for each particle. The heatmaps for the $ddl$ baryon
at $\kappa=0.13700$, \mbox{$\mpi=296\,$MeV} are shown in \autoref{fig:fitting:ddlheatmaps}.

We consider a correlator to have reached single state isolation when the first three windows have
$\chi^2_{\rm dof} < 2.5$ which corresponds to the three leftmost boxes being coloured blue, green
or yellow. We can see this criteria has been fulfilled by all underlying correlation functions by
$t=24$ for the $ddl$ baryon. In determination of the difference we also consider the heatmaps of
the other baryon in question and set $t_{\rm min}$ to the larger of the two earliest times.

As shown in \autoref{sec:quarkmodel:fractionallychargedbaryons} the magnetic polarisability of the
fractionally charged baryons is an order of magnitude smaller than that of physical
baryons. Further, we are interested in the magnetic polarisability difference. The difference is a
very sensitive quantity and extracting meaningful signal is difficult, especially at light quark
masses.

As a result, we will extract two sets of data in this work. One in which the underlying correlation
functions are explicitly checked for single state isolation. This dataset we denote "Corr." to
reflect the correlator level examination of the $\chi^2_{\rm dof}$.  We will also produce a dataset
where this criteria is relaxed and we examine the correlated $\chi^2_{\rm dof}$ of the energy shift
ratio to produce appropriate sampling of the possible fit windows. We denote this dataset
"Ratio".  In this case we set $t_{\rm min} = t_{source}+2$.  We will see in the results
section that the two datasets agree at $1\sigma$.

Having defined values for $\delta E_{\beta_1-\beta_2}(k_d)$ at $k_d=1,\, 2$, we turn our attention to
the Landau term $L(k_d,m_1,m_2)$ of Eq.~\ref{eqn:fitting:fieldstrengthfit}, which depends on the
baryon masses. These masses are also determined through the weighted average approach.
Subtraction of this known term allows the magnetic polarisability difference to be extracted as a
single parameter in the fit of
\begin{equation}
    \frac{\delta E_{\beta_1-\beta_2}(k_d) - L(k_d, m_{B_1},m_{B_2})}{C} = \left(\beta_1-\beta_2\right) \, k_d^2 \, .
\end{equation}
The quality of the fits are similar to those presented in our previous work \cite{kabelitz:octet}.

To determine the uncertainty in the polarisability difference, a jackknife error estimate is
performed.  A second-order jackknife is used to obtain uncertainties on the correlation functions
and correlation function ratios.  Uncertainties for fit values such as the magnetic polarisability
energy shift for each field strength are obtained from individual first-order jackknife
sub-ensembles.  Finally, the fit of the energy shifts as a function of field strength is repeated
on each jackknife sub-ensemble and the error on the ensemble average calculated as the jackknife
error.

\section{Simulation Details}\label{sec:simulationdetails}

The details of the lattice simulation mirror those of our preceding work \cite{kabelitz:octet}. We
reiterate the important details here.

\subsection{Lattice Actions}

\begin{table}[tb]
    \centering
    \caption{Details of the $32^3\times64$ PACS-CS ensembles used in this work. The lattice spacing of each
        ensemble is set using the Sommer scale with $r_0=0.4921(64)(+74)(-2)\,$fm. In all cases
        $\kappa_s^{\rm sea}=0.13640$ and $\kappa_s^{\rm val}=0.13665$~\cite{menaduethesis}. $N_{\rm
                    con}$ describes the number of
        configurations.} \label{tab:simulationdetails:pacsensembles}
    \begin{ruledtabular}
        \begin{tabular}{cccc}
            \noalign{\smallskip}
            $\mpipacs$(MeV) & $\kappa_{u\, d}$ & $a\,$(fm)  & $N_{\rm con}$ \\
            \noalign{\smallskip}\hline\noalign{\smallskip}
            701                   & 0.13700          & 0.1022(15) & 399           \\
            570                   & 0.13727          & 0.1009(15) & 397           \\
            411                   & 0.13754          & 0.0961(13) & 449           \\
            296                   & 0.13770          & 0.0951(13) & 399           \\
        \end{tabular}
    \end{ruledtabular}
\end{table}

The four gauge ensembles used in this work are the four heaviest $2+1$-flavour dynamical gauge
configurations provided by the PACS-CS collaboration~\cite{PACS-CS2008ensensembles} through the
International Lattice Data Grid (ILDG)~\cite{ILDG}. The configurations have a range of degenerate
up and down quark masses while the strange quark mass is fixed.  The Iwasaki gauge action and the
clover fermion action with $C_{SW}=1.715$ are used in generating the configurations.  Details of
the ensembles are summarised in \autoref{tab:simulationdetails:pacsensembles}.

The ensembles were generated in the absence of a background magnetic field.  As a result, the sea
quarks are blind to the magnetic field and the ensembles may be regarded as electro-quenched.  On
the fifth, lightest ensemble, we encounter uncertainties which do not respond to increased
statistics, hinting at an exceptional configuration problem associated with the electro-quenching
of the light sea-quark sector. For this reason, it is omitted here.

Our valence fermion action matches the PACS-CS QCD action and includes the background field
corrected clover term \cite{bignell:2019:cloverpion}. Periodic boundary conditions are used in the
spatial directions.  To avoid signal contamination from the backward propagating states, we use
fixed boundary conditions in the temporal direction. The source is then placed at $t=16$, one
quarter of the total lattice time extent such that one is always well away from the fixed
boundary.

\subsection{Baryon interpolating fields}

The commonly used proton interpolating field in lattice QCD is given by~\cite{Leinweber:1990dv}
\begin{equation}
    \chi_p(x) = \epsilon^{abc} \, \left[ {u^a}^T \, C \, \gamma_5 \, d^{\,b}(x)\right]\,u^c(x) \, ,
\end{equation}
where $C$ is the charge-conjugation matrix. We construct the fractionally charged baryons described
in this work through use of this interpolating field, but provide the appropriate quark propagators
to produce the desired particle. For example, in the construction of the $ddh$ baryon, we provide a
light quark propagator determined with the background field included in place of the doubly
represented up quark and a heavy quark propagator determined at zero-field in place of the singly
represented down quark.

Such interpolating fields, utilised purely with traditional gauge-covariant Gaussian smearing are
ineffective at isolating the baryon ground state in a uniform background
field~\cite{primer2014magnetic,Deshmukh:2018:octet,bignell2020nucleon,Bruckmann:2017pft}. The
uniform background field breaks the spatial symmetry and Landau-mode physics is present at both the
quark and hadronic levels. This must be accommodated.

\subsection{Quark operators}

As detailed in Ref.~\cite{kabelitz:octet}, we utilise asymmetric source and sink operators to
maximise overlap of the lowest energy eigenstates of our baryons in magnetic fields. The operators
we utilise also act to accommodate for the spatial symmetry and Landau-mode issues mentioned above.

Source and sink construction utilises links which are smeared using stout link
smearing~\cite{morningstar:2003linksmearing}. 10 sweeps with isotropic smearing parameter
$\alpha_{\rm stout}=0.1$ are applied to spatially-oriented links.

We construct the quark source using three-dimensional, gauge-invariant Gaussian
smearing~\cite{gusken:1990:smearing}. At all quark masses $\alpha=0.7$ is used while the number of
gauge invariant Gaussian smearing sweeps considered is quark mass dependent. Smearing is tuned to
optimise the onset of early effective-mass plateaus with smaller numbers of sweeps associated with
heavier quark masses. 150 to 350 sweeps are utilised.

The source construction is designed to provide a representation of the QCD interactions with the
intent of isolating the QCD ground state. To encapsulate the quark-level physics of the
electromagnetic and QCD interaction we utilise the eigenmode projection techniques demonstrated in
Ref.~\cite{bignell2020nucleon}, where a comprehensive explanation of the mechanisms may be found.
A succinct summary of the approach is provided in Ref.~\cite{kabelitz:octet}.

The eigenmodes of the two-dimensional lattice Laplacian calculated on the SU(3)$\times$U(1) gauge
links are used to produce a projection operator applied to the propagator at the sink. It is shown
in Ref.~\cite{bignell2020nucleon}, that including too few modes results in a noisy hadron
correlation function in much the same manner as traditional sink smearing. As such, the number of
modes is chosen to be large enough to minimise the noise of the correlation function, but small enough
to retain the focus on the aforementioned low-energy physics. The work of
Ref.~\cite{bignell2020nucleon} found that $n=96$ modes provides balance to these two effects and we
use this here.

\subsection{Hadronic projection}\label{sec:simulationdetails:landauprojection}

\begin{figure*}
    \includegraphics[width=0.99\linewidth]{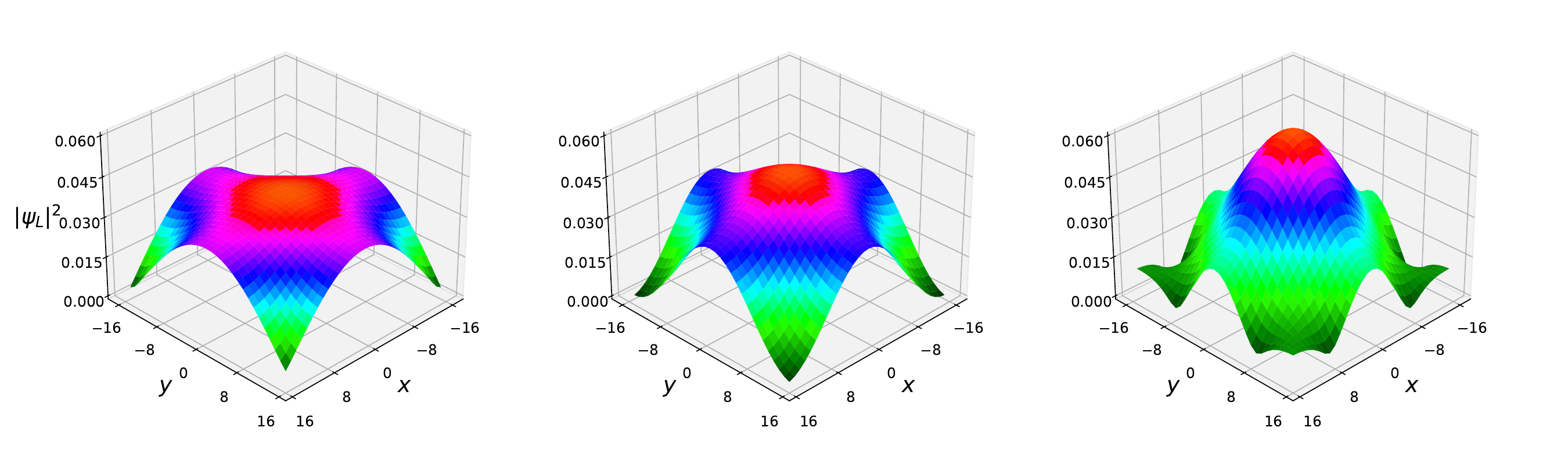}
    \caption{Landau mode probability densities for fractional charge $q_B=-\frac{1}{3}$ and $-\frac{2}{3}$
      baryons at our two lowest field strengths.  Recalling $n=| k_d \,{q_B}/{q_d}|$, the $n=1$
      (left), $n=2$ (centre) and $n=4$ (right) probability densities are illustrated for field
      strengths $|k_d|=1$ and $2$. The degeneracy of each mode is $n$.
      }
    \label{fig:simulationdetails:landaudensities}
\end{figure*}

The inclusion of the background magnetic field induces a change to the wave function of a charged
baryon~\cite{roberts:2010:protonwf}. The quark level electromagnetic physics is highlighted by the
eigenmode projection at the sink. However, we must also ensure that our operator has the
appropriate electromagnetic characteristics on the hadronic level. By projecting final-state
fractionally charged baryons to the state corresponding to the lowest-lying Landau level, we
ensure $n=0$ in Eq.~\ref{eqn:backgroundfield:energyexpansion}. The details of this projection are
presented in Ref.~\cite{kabelitz:octet}.

Due to the colour singlet nature of the baryon, we need only project the eigenmodes of the U(1)
Laplacian rather than the full lattice Laplacian used for the quark-propagator sink. These
eigenmodes carry a degeneracy $n$ in the finite volume of the lattice. For a baryon with charge
$q_B$ the degeneracy for the field strength $k_d$ is
\begin{equation}
    n= \left | k_d \, \frac{q_B}{q_d} \right | \, .
\end{equation}
In this work, we consider fractionally charged baryons with hadronic charge $q_B=-\frac{1}{3}$ and
$-\frac{2}{3}$. At $k_d=\pm 1$ their degeneracies will be 1 and 2 respectively, while at $k_d=\pm
2$ their degeneracies will be 2 and 4 respectively. We utilise a linear combination of the degenerate
eigenmodes which provides optimal overlap with the fermion source.
The probability densities of these hadronic-level projection modes are shown in
\autoref{fig:simulationdetails:landaudensities}.

\subsection{Statistics}

As periodic boundary conditions are used in all four dimensions for the gauge field generation, one
can exploit the associated translational invariance of the gauge fields. A quark source can be
placed at any position on the lattice and then cycled with the gauge field to the standard source position of
$(x,y,z,t)=(1,1,1,N_t/4)$, well away from any fermion action boundary conditions. This enables
additional sampling of the full gauge field.

Further, the two-dimensional nature of the lattice Laplacian operator allows the
eigenmodes for the sink projection to be re-used when the gauge field is cycled solely in the
time direction.
Hence, we increase our statistics on the PACS-CS ensembles by considering four random
spatial sources at $t=16$. The gauge field is then cycled in the temporal
direction by an eighth of the lattice time extent (eight slices in our case) for each
random source. This results in a further increase in statistics by a factor of eight.
Together, random sources and time-direction cycles increase our statistics by a factor of
32.  These multiple samples are binned and averaged as a single configuration estimate in
the error analysis.

\section{Lattice Results}\label{sec:latticeresults}

\begin{figure*}[p]
    {\includegraphics[width=0.8\linewidth]{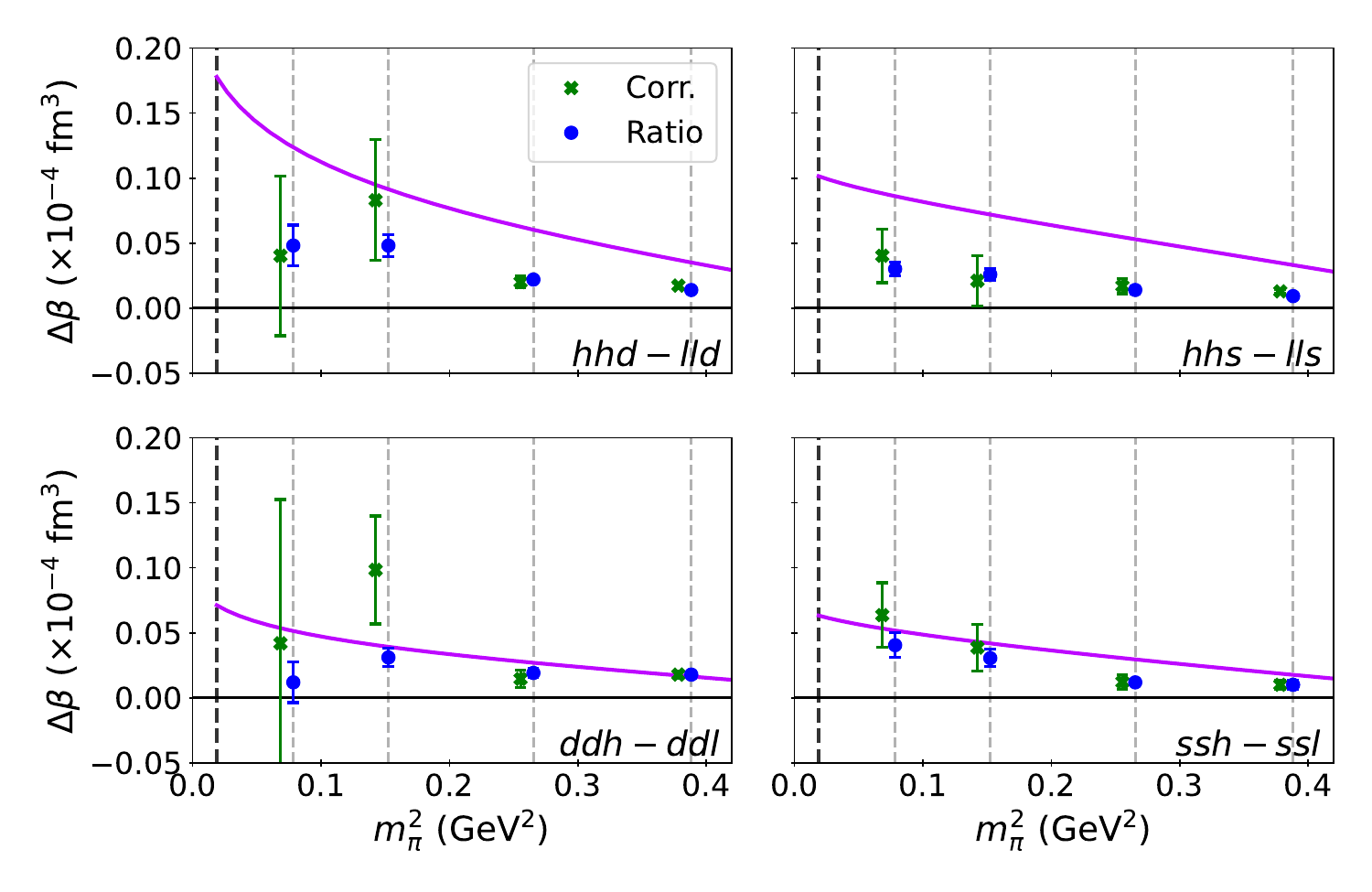}}
    \caption{Indirect environment driven changes in quark-sector contributions to octet baryon
      magnetic polarisabilities, $\Delta \beta$, for singly (upper) and doubly (lower) represented
      quark sectors.  Results are plotted as a function of the pion mass governing the mass of the
      light $d$ and $l$ quarks. The labels in the lower-right of the panels indicate the magnetic
      polarisability difference illustrated.  For example the upper left panel reports $\Delta
      \beta = \beta_{hhd} - \beta_{lld}$. The dataset labels "Ratio" and "Corr." are
      described in the text. The purple curve is the quark model prediction for the environment
      dependence as discussed in Sec.~\ref{sec:QMindirect}.}
    \label{fig:results:environmentaleffects}
\end{figure*}
\begin{figure*}[p]
    {\includegraphics[width=0.8\linewidth]{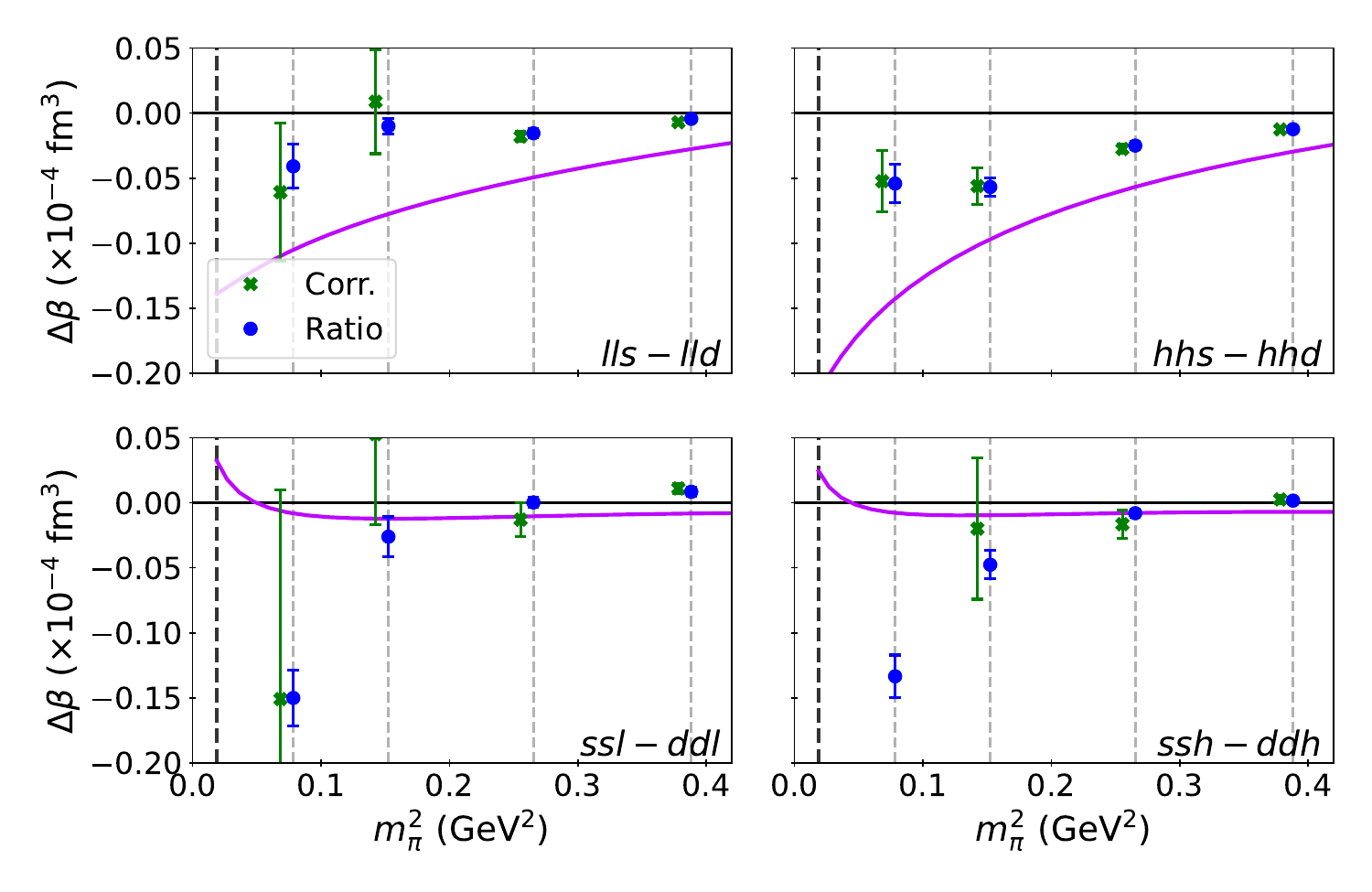}}
    \caption{Direct quark-mass driven changes in quark-sector contributions to octet baryon
      magnetic polarisabilities, $\Delta \beta$, for singly (upper) and doubly (lower) represented
      quark sectors. The presentation of the results is as in
      \autoref{fig:results:environmentaleffects}. Note, the missing data point at the lightest mass
      in the Corr. dataset for $ssh-ddh$ is outside the scale presented at
      $\Delta\beta=-0.41(14)\times 10^{-4}\,$fm$^3$.}
    \label{fig:results:masseffects}
\end{figure*}

In Figs.~\ref{fig:results:environmentaleffects} and \ref{fig:results:masseffects} we show the main
results of this investigation.  The indirect environment driven changes in the magnetic
polarisability are illustrated in Fig.~\ref{fig:results:environmentaleffects}.  From a quark model
perspective this is the simple case where most effects cancel and changes are driven by changes in
the octet-decuplet baryon mass splitting as in Eq.~\ref{eqn:hhd-lld}.  The direct quark mass driven
changes in the magnetic polarisability are illustrated in Fig.~\ref{fig:results:masseffects}.  In
all cases, we have taken the difference with the lighter baryon subtracted from the heavier baryon.

In these figures, two datasets are illustrated. The points labelled "Ratio" are from the
conventional analysis approach where reference to the correlation function ratio is used to
identify fit plateaus.  These results are complemented by the more rigorous results
(offset for clarity) labelled "Corr."which are derived from strict requirements for
single-state isolation in each of the underlying correlation functions involved in the ratio. The
trends observed in both datasets agree.  However, in some cases a loss of signal gives rise to
very large uncertainties in the subtle polarisability changes for the Corr.\ method and we
refer in these cases to the results determined in the Ratio method, as they may be more
representative.

Let us commence with the general trends observed in the results.  As the pion mass increases, all
differences converge toward zero. This is to be expected as the strange quark mass is fixed for all
PACS-CS ensembles.  As the light $d$ and $l$ flavours approach the $s$-quark mass, the
polarisability differences must vanish.

\subsection{Environmental dependence in lattice QCD}
\label{sec:LQCDindirect}

Turning our attention to the environment driven changes in
\autoref{fig:results:environmentaleffects}, we observe a clear environmental effect. We observe
that increasing the mass of the spectator quark results in an increase of the magnetic
polarisability. In \autoref{sec:QMindirect}, we showed that the quark model predicts such an effect
for all four environmental considerations of Eq.~\ref{eqn:4indirect}, For example, consider
Eq.~\ref{eqn:hhd-lld} 
\begin{equation}
    \beta_{hhd} - \beta_{lld} = \frac{4}{9 \, \pi}\left[\frac{\mu_d^2}{m_{\Xi^*} - m_{\Xi}} - \frac{\mu_d^2}{m_{\Delta} - m_N}\right].
\end{equation}
Here, the environment effect is driven by the octet-decuplet mass
splittings which govern the strength of the transition term.
The larger magnetic polarisability of the heavier baryon seen in $\Delta \beta$ is due to the
smaller mass splitting for heavier baryons. In octet (decuplet) baryons the hyperfine attraction
(repulsion) between the quarks is inversely proportional to the product of the constituent quark
masses. As a result, the order of the octet-decuplet mass splittings is $N > \Sigma > \Xi > hhh$.
While the idea that heavier spectator quarks should give rise to larger magnetic polarisabilities
may seem a little counterintuitive at first, the simple quark model provides a compelling
explanation of the complex interactions of lattice QCD.  It is satisfying that these early ideas
founded in perturbation theory are manifest in the full nonperturbative formalism of lattice
quantum field theory.

The quark model's suggestion of a simple effect based only on the octet-decuplet mass splitting
does an impressive job of explaining the dependence, particularly for the doubly represented quark
sector.  The overestimate of the effect in the singly represented sector is interesting.  It has
long been known that the contributions of the singly represented quark flavour to baryon magnetic
moments in lattice QCD is suppressed relative to constituent quark model predictions
\cite{Leinweber:1990dv,Boinepalli:2006xd}. However, this suppression was not observed in the
effective quark moments of the $\gamma N \to \Delta$ electromagnetic transition
\cite{Leinweber:1992pv}.  It would be interesting to revisit these early ideas drawing on modern
lattice QCD techniques.

\subsection{Direct quark mass dependence in lattice QCD}
\label{sec:LQCDdirect}

In the case of the direct quark mass effect, we now know that increasing the mass of the charged
quark of interest results in a decrease in the magnetic polarisability. Intuitively this is a
comfortable result.  However, the quark model presents a
more complex competition of contributions.

Recall the discussion around Eq.~\ref{eqn:lls-lld}
\begin{align}
    \beta_{lls} - \beta_{lld} =
      \frac{4}{9\,\pi}&\left[\frac{\mu_s^2}{m_{\Sigma^*} - m_{\Sigma}} - \frac{\mu_d^2}{m_{\Delta}
        - m_N}\right] \nonumber \\
    - & \left[\frac{\alpha \, q_s^2}{6 \, m_h}\expval{r^2}_h - \frac{\alpha \, q_d^2}{6 \,
        m_l}\expval{r^2}_l\right] \, ,
    \label{eqn:results:ssl-ddl}
\end{align}
in Sec.~\ref{sec:QMdirect}. As the constituent quark moments (Eq.~\ref{eqn:quarkmodel:quarkmoment}) are inversely proportional to quark mass,
increasing the quark mass decreases their contributions, which are squared
in the transition term of the
magnetic polarisability.  In contrast, octet-decuplet mass splittings act in the opposite manner to
increase the magnitude of the transition term, as it did in the environmental case above.  Moreover
the sign of the quark distribution term contributions also act to enhance the magnetic
polarisability.  However, the lattice QCD results now indicate the magnetic polarisability is
suppressed as the quark mass of the active sector is increased, making the decrease in the constituent
quark magnetic moment the dominant effect.

With regard to the singly represented quark sector results in the upper panel of
Fig.~\ref{fig:results:masseffects}, we once again observe an overestimate of the effect in the
quark model.  This is of a magnitude similar to that observed for the singly represented quark
sector in the environmental case.

The tension between these effects increases for the doubly-represented quark sector
where the size of the quark distribution term doubles as in Eq.~\ref{eqn:results:ssl-ddl}, but the sign of the
magnetic polarisability difference in lattice QCD continues to show the same trend.  However, the
quark model predictions are significantly impacted.  This leads to
significant discrepancies for the direct quark mass driven differences in the doubly represented
sector at the lightest quark masses considered where the enhanced quark distribution term $\beta_2$
is now able to dominate the contributions of the mass splitting and quark moments in the quark model.

Clearly, there is present behaviour that the quark model fails to capture.  An increase in the
transition term, $\beta_1$, relative to the distribution term, $\beta_2$, could drive the model
prediction back towards the lattice QCD results.

And given that the singly represented sector overestimates the direct quark mass effect, one needs
to suppress the single quark transition moment.  These effects are in accord with the results of
lattice QCD calculations for octet-baryon magnetic moments
\cite{Leinweber:1990dv,Boinepalli:2006xd}.  Given that the model draws on these for the transition
matrix element, there is qualitative agreement on the manner in which to improve the quark model
further.

\begin{table}[tb]
    \centering
    \caption{Changes in the physical octet baryon magnetic polarisabilities due to an indirect environmental
      effect (top four) and a direct quark mass effect (bottom four). Physical baryon states where
      the effect will come into play are indicated.  The $u$ transitions column is obtained
      by replacing the active $d$ quark with $u$ quark. As there is no octet state composed of three
      heavy quarks, there is no corresponding change.}
    \label{tab:results:physicalcomparison}
    \begin{ruledtabular}
        \begin{tabular}{cccc}
            \noalign{\smallskip}
            Difference & $d$ quark & $u$ quark & Sign\\
            \noalign{\smallskip}\hline\noalign{\smallskip}
            $hhd-lld$ & $\Xi^- - p $ & $\Xi^0 - n $ & $+$\\
            $hhs-lls$ &  & & $+$\\
            $ddh-ddl$ & $\Sigma^- - n $ & $\Sigma^+ - p $& $+$\\
            $ssh-ssl$ &  & & $+$\\
            $lls-lld$ & $\Sigma - p$ & & $-$\\
            $hhs-hhd$ &  & & $-$\\
            $ssl-ddl$ & $\Xi^0 - n$ & &$-$\\
            $ssh-ddh$ &  & & $-$\\
        \end{tabular}
    \end{ruledtabular}
\end{table}

Finally, we connect the implications of these results to the physical octet baryons in
\autoref{tab:results:physicalcomparison}.  Here we list the physical states in which the
differences studied herein can manifest. The first four rows of the table correspond to indirect
environment effects which cause an increase in the magnetic polarisability when the mass of a
spectator quark is increased. The bottom four rows correspond to the direct quark mass effect which
causes a decrease in the magnetic polarisability as the active quark mass is increased.

The most interesting environment effects for understanding the ordering of $n$, $p$, and $\Sigma^+$
baryons are the $hhd-lld$ and $ddh-ddl$ which correspond to the
differences in the contribution of the up quark in $\Xi^0-n$ and $\Sigma^+-p$ due to a change of
environment.  In both cases, the effect enhances the magnetic polarisability of the hyperon
relative to the nucleon. Both
differences also contain a direct mass effect on the other sector given by $ssl-ddl$ and $lls-lld$
respectively. In this case, the magnetic polarisability of the hyperon is suppressed. The opposing
signs of the indirect environment effect and the direct mass effect highlights the complexity in
understanding the magnetic polarisabilities of octet baryons.  We now appreciate that the quark
model underestimates the direct mass effect in the doubly represented sector, leaving the neutron
polarisability higher than it should be.  Thus we have found the origin of the chief discrepancy
depicted in Fig.~\ref{fig:quarkmodel:polarisabilitypredictions}.

\section{Conclusion}\label{sec:conclusion}

Introducing fractionally charged baryons to the background field formalism of lattice QCD, we have
disclosed the individual quark sector contributions to the magnetic polarisabilities of the outer
octet baryons.  We have exposed a direct quark mass effect that acts to suppress the magnetic
polarisability as the quark mass increases.  We have also observed an indirect environmental effect
where the magnetic polarisability is enhanced as the mass of the spectator quark(s) are increased.

To gain an understanding of the underlying dynamics, we explored these direct and indirect effects
in the context of a simple constituent quark model summarised in
Eq.~\ref{eqn:quarkmodel:polarisability}.  The environmental effect appears as a relatively simple
effect and may be explained by the change in the octet-decuplet mass splitting in the transition
term, $\beta_1$. The behaviour of the quark mass dependence is predicted to be more complex in the
model with competition between the intrinsic constituent quark moment and the octet-decuplet baryon
mass splitting in the transition term, $\beta_1$, and further contributions from the quark
distribution term, $\beta_2$.

We now understand that it is the transition term of the polarisability, $\beta_1$ of
Eq.~\ref{eqn:quarkmodel:polarisability}, that contains the physics that describes our lattice QCD
observations.  The direct quark mass effect observed in the lattice results can be associated with
the reduction of the intrinsic constituent quark moment for increasing quark mass, and this moment
is squared in the numerator of $\beta_1$.  The indirect effect observed in the lattice results can
be associated with the reduction of the octet-decuplet mass splitting in the denominator of
$\beta_1$, acting to enhance the magnetic polarisability as the quark mass increases and hyperfine
effects separating the octet and decuplet baryon masses are suppressed.

We were not able to identify an equally important role for the quark distribution term of
$\beta_2$.  In fact its enhancement in the direct quark mass effect in the double-represented quark
sector is problematic and hides the important effect of the transition term.  It is the
distribution term, $\beta_2$, that leaves the neutron magnetic polarisability high relative to the
$p$ and $\Sigma^+$ polarisabilities and the lattice QCD results.

It would be interesting to perform a modern lattice QCD study of the effective quark moments of
octet and decuplet baryons and their electromagnetic transitions to gain more precise insight into
this physics driving direct and indirect environmental effects in baryon magnetic polarisabilities.
Similarly, it will be advantageous to address the exceptional configuration problem encountered at
light quark masses, opening the opportunity to explore the role of chiral physics.  A first step
would be to perform a filtering of the exceptional configurations to establish that the problem is
understood.  Longer term approaches could involve the use of chiral fermion actions to avoid
problematic additive mass renormalisations.  Alternatively, one could employ more efficient fermion
actions and include the electric charges of the dynamical sea quarks such that they may correctly
respond to the presence of a background magnetic field. \\

The results presented herein are founded on the PACS-CS ensembles of Ref.~\cite{PACS-CS2008ensensembles}.
The ensembles are created by the PACS-CS collaboration, Aoki {\it et al.} and are available from \url{https://www2.jldg.org/}.
The ensembles in question are part of the April 2009 release.

\begin{acknowledgments}
We thank the PACS-CS Collaboration for making their $2+1$ flavour configurations available and the
ongoing support of the International Lattice Data Grid (ILDG).  Baryon correlation functions were
constructed using the \texttt{COLA} software library, developed at the University of Adelaide~\cite{COLA}.
WK was supported by the Pawsey Supercomputing Centre through the Pawsey Centre for Extreme Scale
Readiness (PaCER) program. This work was supported with supercomputing resources provided by the
Phoenix HPC service at the University of Adelaide. This research was undertaken with the assistance
of resources from the National Computational Infrastructure (NCI), which is supported by the
Australian Government. This research is supported by Australian Research Council through Grants
No.~DP190102215 and No.~DP210103706.
\end{acknowledgments}

\bibliography{main}

\end{document}